\documentclass[manuscript]{aastex}

\slugcomment{Accepted for publication in The Astrophysical Journal}

\shorttitle{The Origin of O{\sc vi} Absorbers}
\shortauthors{Heckman et al.}

\begin{document}


\title{On the Physical Origin of O{\sc vi} Absorption-Line Systems}


\author{T.M. Heckman\altaffilmark{1}, C.A. Norman\altaffilmark{2},
D.K. Strickland\altaffilmark{3},
\& K. R. Sembach\altaffilmark{1}}
\affil{Department of Physics \& Astronomy, Johns Hopkins University,
Baltimore, MD 21218}


\altaffiltext{1}{Guest Investigators on the NASA-CNES-CSA Far
Ultraviolet
Spectroscopic Explorer. FUSE is operated for NASA by the Johns Hopkins
University under NASA contract NAS5-32985.}
\altaffiltext{2}{Space Telescope Science Institute, Baltimore, MD 21218}
\altaffiltext{3}{Chandra Fellow}


\begin{abstract}
We present a unified analysis of the O{\sc vi} absorption-lines
seen in the disk and halo of the Milky Way, high velocity clouds,
the Magellanic Clouds, starburst galaxies, and the intergalactic medium.
We show that these disparate systems 
define a simple relationship between the O{\sc vi}
column density and absorption-line
width that is independent
of the Oxygen abundance over the range O/H $\sim$ 10\% to twice solar.
We show that this relation is exactly that
predicted theoretically as a radiatively cooling flow of hot gas passes 
through the coronal temperature regime - independent of its
density or metallicity (for O/H $\gtrsim$
0.1 solar). Since most of the
intergalactic O{\sc vi} clouds obey this relation, we infer that they can not
have metallicities less than a few percent solar.
In order to be able to cool radiatively in less than a Hubble time,
the intergalactic clouds must be smaller than $\sim$1 Mpc in size.
The implied global
heating rate of the warm/hot IGM is consistent with available sources.
We show that the cooling column densities for the
O{\sc iv}, O{\sc v}, Ne{\sc v}, and Ne{\sc vi} ions are comparable
to those seen in O{\sc vi}. This is also true for
the Li-like ions Ne{\sc viii}, Mg{\sc x}, and Si{\sc xii} (if
the gas is cooling from $T \gtrsim 10^6$ K).
All these ions have strong resonance lines in the
extreme-ultraviolet spectral range, and would be accessible to $FUSE$
at $z \gtrsim$ 0.2 to 0.8. 
We also show that
the Li-like ions can be used
to probe radiatively cooling gas at temperatures an order-of-magnitude
higher than where their ionic fraction peaks.
We calculate that the H-like (He-like)
O, Ne, Mg, Si, and S
ions have cooling columns of $\sim10^{17}$ (few$\times10^{16}$)
cm$^{-2}$. The
properties of the O{\sc vii}, O{\sc viii}, and Ne{\sc ix} 
X-ray absorption-lines
towards PKS 2155-304 may be consistent with a scenario of radiatively
cooling gas in the Galactic disk or halo.
\end{abstract}


\keywords{intergalactic medium - galaxies: ISM - galaxies: halos
- Galaxy: halo - ISM: general}


\section{Introduction}

Diffuse gas in the temperature range of $\sim 10^5$ to $10^6$ K
(sometimes called ``coronal gas'')
is of considerable interest in many astrophysical
contexts. The peak of the radiative cooling curve for plasmas
lies at the low temperature
end of this range for metallicities greater than a few percent solar
(e.g., Sutherland \& Dopita 1993), and since the cooling curve
declines with increasing temperature in the coronal regime, this gas will be
thermally unstable.

In the interstellar medium (ISM)
of our own galaxy, coronal gas is created as a consequence
of mechanical heating
associated with massive stars - supernovae and stellar winds
(e.g., Cox \& Smith 1974; Weaver et al. 1977).
It is believed that coronal gas traces the interface between hotter
(X-ray-emitting) gas and cooler clouds. The rate of mass exchange
between the hot and cool phases and the related issue of the
fraction of the volume of the disk occupied by the hot gas
are crucial components of models of the global structure of the ISM
(e.g. McKee \& Ostriker 1977).

Since the sound speed in coronal gas exceeds the typical
velocity dispersion in the disk of a star-forming galaxy,
coronal gas created in the disk can - in the absence of rapid
radiative cooling - flow upward to large
vertical scale-heights and occupy the lower halo of the galaxy
(Shapiro \& Field 1976; Bregman 1980; Norman \& Ikeuchi 1989).
On larger scales, since the virial temperature for typical galaxies and
small galaxy groups lies
near the upper limit of the coronal regime, this gas may be an important
component of the large-scale gaseous halos of galaxies and of the
intra-group medium (e.g., Norman \& Silk 1979;
Verner, Tytler, \& Barthel 1994; Mulchaey et al. 1996).

The most significant repository of coronal gas is the intergalactic 
medium (IGM). Gas and stars inside galaxies
account for only 10 to 20\% of the baryonic content of the universe
(e.g., Fukugita, Hogan, \& Peebles 1998). Numerical cosmological simulations
imply that the bulk of the ``missing baryons'' lies outside galaxies, groups,
or clusters (e.g., Cen \& Ostriker 1999; Dav\'e et al. 2001). They also imply
that shocks generated by the formation of large-scale structure
(and possibly by galactic winds) have heated a substantial fraction
of these baryons to temperatures in the coronal regime (the Warm-Hot
Intergalactic Medium, or WHIM).

The $Far~Ultraviolet~Spectroscopic~Explorer$ mission ($FUSE$ - Moos et al. 
2000)
has opened the coronal-phase in the ISM and IGM in the low-z universe
to direct observation via the
O{\sc vi}$\lambda$$\lambda$1032,1038 doublet. Oxygen
is the most cosmically-abundant metal. For
collisional ionization equilibrium (CIE), O{\sc vi} is a significant species
for $T \sim (2- 5)\times 10^5$ K. 
It is also a byproduct
(and hence an important tracer) of hotter radiatively-cooling gas 
(e.g., Edgar \&
Chevalier 1986).

To date, $FUSE$ has detected the O{\sc vi}
line in absorption in the disk
(Jenkins, Bowen, \& Sembach 2001) and halo (Savage et al. 2000) 
of the Milky Way,
in the periphery of high velocity clouds (Sembach et al. 2000), in the
interstellar medium of the Large and Small Magellanic Clouds
(Hoopes et al. 2002; Howk et al. 2002), in several starburst galaxies 
(Heckman et al. 2001a,b; Martin et al. 2002), and in the IGM 
(Oegerle et al. 2000; Sembach et al. 2001; Savage et al. 2002).

In each case, the authors have discussed the origin and physical state
of the absorbing gas in the particular astrophysical context of
their investigation. Our goal in the present paper is to consider
the class of O{\sc vi} absorbers more generally. In particular, we will show
below that {\it all} of the above classes of O{\sc vi} absorbers obey the 
simple relationship between the O{\sc vi} column density and line 
width that is predicted as radiatively cooling hot gas passes
through the coronal regime.

\section{Data Analysis \& Results}

\subsection{Data Analysis}

In this investigation we make use of existing O{\sc vi} data in the 
literature and in a few cases supplement this with additional data that 
are currently being analyzed.  The primary quantities pertinent to our
investigation are the column density of O{\sc vi} and the velocity width
(line profile second moment) of the O{\sc vi} absorption in different 
astrophysical regions. 
 We consider data that have been published for the following
locations:
Galactic disk (Jenkins 1978; $Copernicus$ data), Galactic halo and 
high velocity clouds (Savage et al. 2000, 2002; Sembach et al. 2000, 2001, 
2002; Richter et al. 2001a, 2001b; Heckman et al. 2001a,b), LMC (Howk et al. 
2002),
SMC (Hoopes et al. 2002), starbursts (Heckman et al. 2001a, 2001b), and
the IGM ([PG0953+415: Tripp \& Savage 2000 -- STIS; Savage et al. 2002]; 
[3C\,273: Sembach et al. 2001]; [H\,1821+643: Tripp, Savage, \& Jenkins 
2000 -- STIS; Oegerle et al. 2000]).

Unless specified otherwise, these results are based
upon $FUSE$ data.  In most cases, information is available for only the 
O{\sc vi}$\lambda$1032 line since the 1038\,\AA\ line is blended 
with other ISM absorption features along the sight lines.  Column densities 
were obtained by integrating the apparent optical depths of the 1032\,\AA\
lines, and the line widths were calculated as b$/\sqrt{2} =
\sigma = \sqrt{\int{(v-\bar{v})^2 \tau_a(v)dv}/\int{\tau_a(v)dv}}$ over 
the same velocity ranges used to calculate the column densities.

We retrieved the $FUSE$ data for I\,Zw\,18, Mrk~54, and NGC~7673
from the $FUSE$ archive to supplement the starburst 
sample obtained by Heckman et al.
(2001a, 2001b).  We treated the spectra in a manner similar to that of the 
data in our previous studies. For I\,Zw\,18, we find  $\log N_{OVI} < 13.6~ 
(3\sigma)$ in the $\pm100$ km~s$^{-1}$ velocity range centered on the systemic
velocity $v_\odot = +750$ km~s$^{-1}$.  This limit is roughly an order of 
magnitude smaller than the O{\sc vi} columns found for other starbursts.
For Mrk~54 and NGC~7673 we find $\log N_{OVI}$ = 14.9 and 15.0 and
b = 150 and 180 km~s$^{-1}$ respectively.

\subsection{Results \& Possible Selection Effects}

Our principal result is shown in Figure 1, where we have plotted the
O{\sc vi} column density
($N_{OVI}$) and line widths (b-values) for the various
systems listed above.
 
Taken together, the various O{\sc vi} systems define a clear
relationship that spans over two-orders-of magnitude in
column density and over one order-of-magnitude in line width. The
relation is a tight one, with a scatter of only $\sim\pm$0.1 dex in log b
at a given value of $\log N_{OVI}$. Above log b $\sim$ 1.6, the
column density increases linearly with line width, while at lower
line widths the relation rolls over and steepens.
Different types of O{\sc vi} systems preferentially
populate different regions in parameter space, but with
considerable overlap. The Galactic disk,
high velocity cloud, and IGM systems tend to
have the smallest column densities and line widths,
and the starbursts have the largest. The Magellanic Cloud and Milky Way halo
systems are intermediate, overlapping well with the low-b, low-column
starburst systems.
We will discuss the physical interpretation of this correlation in detail
in section 3 below. We will show there that the observed correlation
is well fit by a simple model {\it with essentially no adjustable parameters.}

Before proceeding to this discussion, it is important to establish
that the correlation in Figure 1 is not due to an observational selection
effect. More specifically, for lines in the linear portion of the
curve-of-growth, the ratio of the column density and line width
is just proportional to the optical depth of the absorber. Broad lines
associated with gas of small column density will correspond to small
optical depths. Such shallow lines might elude detection, leading
to an absence of detected lines in the lower right corner of Figure 1.
The dotted line in the lower right corner of Figure~1 indicates the expected
behavior for a line 
with $N/\sigma = 7.5\times10^{11}$ cm$^{-2}$ km~s$^{-1}$, the approximate
$2\sigma$ detection threshold for the $FUSE$ O{\sc vi} $\lambda1032$ 
data considered.

We believe this selection effect is not responsible for the correlation
in Figure 1 for several reasons. First, it would not account for the absence
of points in the upper left corner of Figure 1. Lines produced by
gas with a large column density and small line width
would be optically-thick and black at line center. These lines
would be easy to detect and recognize, if such gas were commonplace.

Second, O{\sc vi}
absorption is {\it already} seen by $FUSE$ in essentially all the sight lines
through the Galactic disk and halo, near high velocity clouds,
in the Magellanic Clouds,
and in starbursts (except when the relevant wavelength
range was contaminated by other strong lines). Thus, only in the case
of the intergalactic clouds (whose redshifts are not known
{\it a priori}) could selection against weak, broad lines come into play.

In fact, larger samples of O{\sc vi}
absorption systems in the Galactic halo (Savage et al. 2002) and in
high velocity clouds (Sembach et al. 2002) further strengthen the relationship
shown in the figure. These larger samples are particularly significant because 
they are obtained using the light of $\sim100$ quasars and active galactic
nuclei as background sources.
These objects have very smooth ultraviolet continua,
thus minimizing continuum placement uncertainties for any intervening O{\sc vi}
absorption features. Numerous broad features exist within those samples, but
they typically follow the trend shown in Figure 1.
The best example of a broad, weak
absorption wing is given by Sembach et al. (2001) for the Galactic O{\sc vi}
absorption in the direction of 3C\,273 at velocities
$v_{LSR} > 100$ km\,s$^{-1}$.
Such absorption wings are relatively uncommon and fall well above the $FUSE$
detection threshold indicated in the figure.

To further illustrate this, we
show in Figure 2  spectra of a broad (log~b = 2.25) redshifted O{\sc vi}
synthetic absorption-line added to a typical $FUSE$ spectrum. In the
the three cases plotted, the line has a central residual
intensity of 85\% (corresponding to the
$FUSE$ detection threshold in Figure 1), 28\% (roughly corresponding to the
solid line passing through the data points in Figure 1), and
64\% (intermediate between the previous two). The signal-to-noise in the
the continuum is representative of the data plotted in Figure 1.
It seems clear that the lack of broad, weak O{\sc vi} lines in Figure 1
can not be purely due to detectability issues.

\section{Discussion}

\subsection{The Physics of Radiatively Cooling Gas}

Consider the generic situation in which
gas cools radiatively from an initial temperature $T$ and density $n$
(e.g., Edgar \& Chevalier 1986).
The total column density of cooling gas
is given by $N_{cool} = \dot{N} t_{cool}$, where
$t_{cool}$ is the cooling time and
$\dot{N}$ is the rate of cooling per unit area (cm$^{-2}$ s$^{-1}$). The ratio
$\dot{N}/n$ has dimensions of velocity - an expression
of mass conservation in the cooling gas.
Thus, the total cooling column density can be
written as $N_{cool} = n t_{cool} v_{cool}$. In the case of radiative cooling
of gas with temperature $T$ and metallicity $Z$,
$t_{cool} = 3 kT/n \Lambda (T,Z)$, where
$\Lambda$$(T,Z)$
is the cooling function. Thus $N_{cool} = 3 kT v_{cool}/\Lambda (T,Z)$.

The characteristic velocity $v_{cool}$ in the above expression has
slightly different physical meanings in different contexts. For example,
in the case of gas cooling behind a shock the appropriate velocity
is the postshock flow velocity in the O{\sc vi} zone, which is related to
the cooling length $l_{cool} \sim v_{cool} t_{cool}$
(hence $N_{cool} = n l_{cool}$). In gas collapsing in a cooling instability
the characteristic size of the cooling region is $l_{cool} \sim c_s t_{cool}$,
where $c_s$ is the isothermal sound speed $(P/\rho)^{1/2}$. Thus,
the appropriate velocity to use in this case is the sound speed.

The expression $N_{cool} = 3 kT v_{cool}/\Lambda (T,Z)$
shows explicitly that $N_{cool}$ is
independent of density for radiative cooling.
Moreover, at coronal temperatures radiative
cooling is dominated by resonance lines due to metals 
(primarily Oxygen for log $T$ = 5.0 to 5.6,
Neon for log $T$ = 5.6 to 5.9, and Iron at higher temperatures). Thus,
$N_{cool} \propto \Lambda^{-1}.
\propto (Z)^{-1}$.
\footnote{The cooling functions computed by Sutherland \& Dopita (1993)
show that for coronal gas in CIE,
$\Lambda \propto Z^{-1}$  is valid for Oxygen and Neon abundances
$Z_{O,Ne} >$ few \% solar. For the
non-equilibrium cases they consider, the proportionality holds
for $Z_{O,Ne} >$ 10\% solar. Note that their 
models are parameterized by the Fe abundance, and they
assume that [O/Fe] = [Ne/Fe] = +0.5 dex 
for [Fe/H] $\leq$ -1.0.} 
Taking the ionic fraction of O{\sc vi} as $f_{OVI}(T)$, we
have $N_{OVI} = 3 kT v_{cool} (O/H)_{\odot} Z f_{OVI}/\Lambda$. 
This column should properly
be calculated as a path-integral through the cooling column of gas. However,
since $f_{OVI}$ is appreciable over a rather narrow range in temperature
centered at $\sim10^{5.5}$ K ($\equiv T_{OVI}$) for CIE, $N_{OVI}$
can be approximated by evaluating the above expression for $T = T_{OVI}$.
This yields $N_{OVI} = 3 \times 10^{14}$ cm$^{-2}$ for
$v_{cool} = 10^2$ km s$^{-1}$.
\footnote{We note that the above estimate assumes that the gas is optically
thin to its cooling radiation. In CIE at $T = 3 \times 10^5$ K the most
important coolants are (in rank order) the \ion{O}{5}$\lambda$629, 
\ion{O}{5}$\lambda$1218, and \ion{O}{6}$\lambda$$\lambda$1032,1038 lines which
together
carry 80\% of the total cooling (using the MEKAL plasma code). Based
on the observed \ion{O}{6} columns and the implied \ion{O}{5} columns
for cooling gas (Table 1), we find that the effective cooling rate would be
reduced by
0.0 to 0.2 dex for the systems in Figure 1.}
 
This simple estimate is confirmed by more detailed
calculations of radiatively cooling gas.
Edgar \& Chevalier (1986) have computed O{\sc vi} columns for the generic
case in which hot gas cools from $T = 10^6$ K.
They consider isobaric, isochoric, and intermediate cases
and find $N_{OVI} \sim 4 \times 10^{14} (v_{cool}/10^2$ km s$^{-1})$ cm$^{-2}$.
Calculations of collisionally-ionized gas behind a high-speed
radiative shock (Dopita \& Sutherland 1996)
give values for $N_{OVI} = (2-3) \times 10^{14}$
cm$^{-2}$ for shock speeds ranging from 200 to 500 km s$^{-1}$
(postshock flow speeds of $v =$ 50 to 125 km s$^{-1}$).
\footnote{For $v_{shock} >$ 300 km s$^{-1}$, higher O{\sc vi}
columns can be produced
via photoionization of 
preshock gas by EUV radiation from the shocked gas, if this
gas is optically-thick to the radiation.}

Begelman \& Fabian (1990) have considered ``turbulent
mixing layers'' that arise at the interface between a flow of
hot gas past cool gas. The mixing of hot and cool gas creates
gas at intermediate temperatures which then cools radiatively. Their model
implies $N_{cool} = 5 \eta kT v_t/\Lambda(T)$, where 
$v_t$ is the turbulent velocity and  $\eta$ is a dimensionless efficiency
factor which they argue is of order unity. 
Their equation (4) 
leads to a predicted O{\sc vi} column of $\sim 3\times 10^{14} \eta$
cm$^{-2}$ for a turbulent velocity of 100 km~s$^{-1}$ and a mixing layer
temperature of $10^{5.5}$ K (=$T_{OVI}$), where we have
used $\Lambda$$(T,Z)$ and $f_{OVI}(T)$
for CIE from Sutherland \& Dopita (1993).

We can also cast results for conductively heated gas in the same terms.
For the case of saturated conduction (Cowie \& McKee 1977), the heat flux
is given by $q_{sat} = 5 \phi c_{s} P$
where $c_{s}$ is the isothermal
sound speed, $P$ is the gas pressure,
and $\phi$ is a dimensionless
constant of order unity. If we balance this heat flux
by radiative cooling we get $q_{sat} = n^2 \Lambda D$,
where $D$ is the thickness
of the conductively-heated interface. 
Combining these relations, we have
$N_{cool} = nD = 5 \phi c_s P/n \Lambda = 10 \phi c_s kT/\Lambda$. Apart from
constants of order unity, we recover the same expression as that
above for $N_{cool}$ (where $v_{cool} \sim c_s$).
\footnote {In the case of classical (unsaturated) conduction,
similar arguments give $N_{cool} = [\kappa(T) T/\Lambda(T,Z)]^{1/2}$, where
$\kappa$ is the thermal conductivity (Cowie \& McKee 1977). While $N_{cool}$
is still independent of $n$, $N_{OVI}$
is no longer independent of metallicity, but will scale as $Z^{1/2}$.
For solar metallicity
and CIE at $T = 10^{5.5} K$
this predicts $N_{OVI} \sim 3 \times 10^{13}$ cm$^{-2}$.}

A robust conclusion emerges from the above.
{\it The characteristic O{\sc vi} column density in radiatively
cooling coronal gas is essentially independent of density, metallicity,
and the heating mechanism. It
depends only on the value for $v_{cool}$, the characteristic cooling
flow speed.}

\subsection{Comparison to the Data}

In order to compare the predictions of the model described above
to the results in Figure 1 we need to relate the observed line
width $b_{obs}$ to the cooling flow velocity $v_{cool}$.

In the absence of a flow, the O{\sc vi} lines will have an irreducable
intrinsic width due purely to thermal broadening: $b_{therm} = (2kT/m_O)^{1/2}
= 10 (32)$ km s$^{-1}$ for $T = 10^5~(10^6)$ K.
The observed line
broadening is due to the combination of the thermal broadening
and that produced by the cooling
flow: b$_{obs} = \sqrt{b_{thermal}^2+b_{cool}^2}$.
We identify $b_{cool}$ with the cooling flow velocity $v_{cool}$ that sets 
the O{\sc vi} column density. We have therefore computed the relation 
between $N_{OVI}$ and b$_{obs}$ expected
for radiatively cooling gas with 
b$_{thermal}$ corresponding to $T_{OVI} = 10^5$ and $10^6$ K
(chosen as representative bounds for the collisionally ionized gas).

The fit of this relationship to the data is excellent. It is important
to emphasize that this model has essentially no tunable parameters
(apart from $T_{OVI}$, whose range of allowed values is robustly set by
simple physics). In particular, this model naturally explains the linear
proportionality between b$_{obs}$ and $N_{OVI}$ for large b$_{obs}$
(b$_{obs} >>$ b$_{therm}$), the value of the ratio $N_{OVI}$/b$_{obs}$
in the ``linear regime'',
and the ``turnover'' in the relationship
at small b$_{obs}$ (when b$_{obs} \gtrsim $ b$_{therm}$).
\footnote
{A multiplicity of velocity components at different velocities may also
affect the observed line widths.  If identical components have the same
central velocity, they would follow the linear relationship shown in the
figure (neglecting thermal broadening).  If, however, the components are
separated by an amount $\Delta$$v$, there will be an additional broadening
that will cause deviations from this linear relationship.  We have shown
the curves for a set of three identical components having b = 17 km~s$^{-1}$
and symmetrical separations of
$\Delta$$v$ = $\pm$ 0, 10, 20, and 30 km~s$^{-1}$ about the central component
(thin solid lines).  Larger values of $\Delta$$v$ move points further to the
right in this plot.}

A crucial prediction of the radiatively cooling scenario
is that the relationship
between O{\sc vi} column density and line width should be independent
of the metallicity of the cooling gas, provided that the metallicity
is greater than $\sim$ 0.1 solar. This prediction is directly confirmed
by Figure 1, which includes O{\sc vi} systems that span a metallicity range from
$\sim$10\% solar (SMC sightlines and the starburst NGC1705) through
$\sim$solar (Milky Way disk and halo and the starburst NGC3310) to
several times solar (the M83 starburst nucleus).

Our only probe of the
regime where metallicity effects should become significant is
the extremely metal-poor starburst  I~Zw~18 (O/H = 2\%).
Interestingly, we have only
an upper limit to the O{\sc vi} absorption in this case:
$N_{OVI}$/b$_{obs} \leq
5 \times 10^{11}$ cm$^{-2}$ per km s$^{-1}$. This is about an 
order-of-magnitude below the average ratio seen for the other starbursts
in Figure 1. 
Since it is likely that the metallicity
in the bulk of the ISM of I~Zw~18 is even lower than the 2\% solar
value in the HII regions (e.g. Thuan et al. 2002), the absence of detectable
O{\sc vi} absorption in this galaxy is consistent with our model.

The correlation in Figure 1 is clearly a sequence along which the dynamical
state of the O{\sc vi} absorbers changes significantly.
Dividing the measured b-values by the speed-of-sound at $T_{OVI}$
allows us to roughly characterize the
Mach number of the flow ($\cal M$). This provides a simple 
(but physically instructive) summary of the observational data.
We see that the Galactic disk, high velocity cloud, and intergalactic O{\sc vi}
systems are quiescent, highly subsonic flows (possibly
arising in conductive interfaces
between hotter and cooler gas). The Milky Way halo, Magellanic Cloud,
and several of the starburst systems appear to be trans-sonic flows
($\cal M \sim$ 1). This is consistent with models for
fountain/chimney flows
(e.g. Shapiro \& Field
1976; Norman \& Ikeuchi 1989), or dynamically
young galactic outflows in the blow-out phase (e.g. Heckman et al. 2001a).
Finally, the most extreme starburst systems (those in NGC 3310, NGC 7673,
and Mrk 54) show supersonic flows with Mach numbers of several. In these
cases the O{\sc vi} absorption is significantly blueshifted relative to the
galaxy systemic velocity, and we are probably observing
well-developed galactic winds (e.g. Heckman 2001). 

\section{Implications}

\subsection{The Nature of the Intergalactic O{\sc vi} Clouds}

We have argued above that the striking correlation in Figure 1 arises naturally
from the simple physics of radiatively cooling hot gas.
In principle O{\sc vi} can also be produced by
the photoionization of relatively cool gas by a suitably hard ionizing source.
The only physically plausible such source in the present context is
the metagalactic ionizing background due to QSOs. This mechanism can
be ruled out for most of the classes of O{\sc vi} absorbers in 
Figure 1 (e.g., Sembach et al. 2000; Savage et al. 2000) since
the feeble intensity of the metagalactic background implies that implausibly
large and low-density clouds are required to produce the observed O{\sc vi}
columns. A local photoionizing source is implausible in these cases, since
normal stars produce
insignificant amounts EUV radiation with $h\nu \geq$ 114 eV 
($\chi$$_{OVI}$).

In principle, either collisional ionization or
photoionization by the metagalactic background
could produce a significant
fraction of the {\it intergalactic} O{\sc vi} clouds (Tripp \& Savage 
2000; Tripp et al. 2001;
Cen et al. 2001). We would argue that
Figure 1 strongly suggests that most intergalactic O{\sc vi} clouds 
(like the other
O{\sc vi} absorption-line systems) are collisionally-ionized, 
radiatively-cooling gas. In fact, six out of the eight IGM clouds
in Figure 1 lie within the same range in parameter space as the
high velocity cloud and Milky Way disk systems. 
{\it There is no plausible reason why photoionization should naturally
place the intergalactic clouds directly on top of the correlation
between O{\sc vi} column density and line width defined by collisionally
ionized coronal gas in the other systems.} 
\footnote{We note that two of the IGM points do lie well above the trend of the
rest of the points shown in Figure~1.  These are the $z=0.06807$ and
$z=0.14232$ absorbers toward PG\,0953+415. The absorption produced by
\ion{O}{6} and other ions in both absorbers can be described
self-consistently by photoionization models of dilute gas if absorption
arises over path lengths of $\sim100$ kpc (see Savage et al. 2002).}

The total ionized gas column density in a photoionized cloud 
depends only on the ionization parameter ($U$, the ratio of photon
to electron density) and the cloud's optical depth in the Lyman continuum:
$N_{+} \sim (1 - e^{-\tau})Uc/\alpha$, where $\alpha$ is the recombination
coefficient. The predicted O{\sc vi} column is just
$N_{OVI} = 8.5\times10^{-4} Z f_{OVI} N_{+}$. In the intergalactic clouds
the relative strength of the O{\sc vi} line requires $U \gtrsim 10^{-1}$ and
$f_{OVI}$ to be near its maximum value (of order $10^{-1}$) 
(Tripp \& Savage 2000). The predicted O{\sc vi} column
is $N_{OVI} \sim 10^{18} Z (1 - e^{-\tau})$ cm$^{-2}$ for
$U \sim f_{OVI} \sim 0.1$. Thus, photoionization
does not lead to any ``natural'' value
for $N_{OVI}$, nor does it imply any simple proportionality between $N_{OVI}$
and line width (Figure 1).

For our hypothesis of radiative cooling of hot gas to be correct,
it follows first that the intergalactic O{\sc vi}
clouds must have metallicities greater than a few \% solar. This
is consistent with theoretical models for the WHIM in
which it has been chemically-enriched 
by the outflow of metals from galaxies
(Aguirre et al. 2001; Cen et al. 2001).

Our hypothesis also requires the clouds to have
radiative cooling times less than the Hubble time. This leads directly to
a constraint on the sizes of the O{\sc vi} clouds. 
The cooling time for gas in CIE
at $T_{OVI} = 10^{5.5}$ K is 
$t_{cool} = 4.7 \times 10^{3} n^{-1} Z^{-1}$ years (Sutherland
\& Dopita 1993). The O{\sc vi}
column for region with a size $D_{OVI}$ is given by 
$N_{OVI} \sim 8.5\times10^{-4} Z n D_{OVI} f_{OVI}$. Combining these
relations and evaluating $f_{OVI}$ at $T_{OVI}$, we have $t_{cool} = 
2 \times 10^{10} (D_{OVI}/$Mpc) years. Thus, independent of the metallicity,
the requirement $t_{cool} \leq t_{Hubble}$ requires $D_{OVI} \lesssim$ 1~Mpc.
We note parenthetically that the ratio of the radiative cooling
and sound-crossing time will be of-order unity in these clouds, so
the cooling would be intermediate between the isobaric and isochoric cases
considered by Edgar \& Chevalier (1986) and in Table 1.

The cooling constraint can also be cast in terms of density.
For $H_0$ = 70 km s$^{-1}$ Mpc$^{-1}$ and $\Omega_B$ = 0.04,
$t_{cool} \leq t_{Hubble}$ requires that the O{\sc vi} clouds correspond
to mean overdensities $\rho/\bar{\rho} \geq 1.3 Z^{-1}$. Thus, for
$Z$ = 0.03 to 0.3 in the O{\sc vi} clouds (Aguirre et al. 2001; Cen et al. 
2001)
only moderate overdensities are needed. This agrees with the computations
reported by Cen et al. (2001) in which the bulk of the 
hot O{\sc vi} absorbing clouds have overdensities of 20 to 40.

If the O{\sc vi} clouds are cooling on cosmologically-short timescales,
they require a suitable heating source.
In the numerical simulations
by Cen et al. (2001) and Dav\'e et al. (2001) this heating
is provided by shocks and compression associated
with the formation of large-scale structure.
At least on the basis of simple energetic considerations, 
supernova-driven galactic winds could also heat this material.
Tripp, Savage, \& Jenkins (2000) have estimated that the global mass
density of intergalactic O{\sc vi} clouds is 
$\Omega_{OVI} = 4\times10^{-4} Z^{-1}$ or 
$\rho_{OVI} = 6\times10^7 Z^{-1} M_{\odot}$ Mpc$^{-3}$. For a temperature
$T_{OVI} = 10^{5.5}$ K, the corresponding thermal energy density
is $1.4\times10^{55} Z^{-1}$ ergs Mpc$^{-3}$. We can compare that
to the time-integrated energy density due to supernovae. The galaxy
luminosity density measured by Blanton et al. (2001) implies
a mean stellar mass density of $6\times10^8 M_{\odot}$ Mpc$^{-3}$
(Fukugita, Hogan, \& Peebles 1999). Taking a yield of one supernova
per $10^2 M_{\odot}$ implies that supernovae have supplied
$6\times10^{57}$ ergs Mpc$^{-3}$. For $Z$ = 0.03 to 0.3, the
total supernova energy is $\sim$ one to two orders-of-magnitude 
larger than the thermal
energy of the O{\sc vi} clouds. Cen et al.(2001) estimate that
the O{\sc vi} clouds comprise of order 10\% of the WHIM. Given the apparently
high efficiency with which starburst-driven winds transport supernova
kinetic energy (Heckman 2001), we conclude they could be
a significant heating source in the intergalactic O{\sc vi} clouds.

\subsection{Predictions for other Coronal Ions}

Our hypothesis that the observed O{\sc vi} originates in radiatively cooling
hot gas makes an interesting prediction about the detection
of both the hotter gas that is the precursor to the O{\sc vi}
and the cooler gas that is its byproduct. The columns in
the hotter gas are substantial, essentially because the cooling
times are long. We have $N_{cool} = 3 kT v_{cool}/\Lambda (T)$. Assuming that
$v_{cool} \propto c_s$, and approximating $\Lambda (T) \propto T^{-0.8}$ from
$\log T =$ 5.5 to 7.0 (Sutherland \& Dopita 1993), yields $N_{cool} \propto
T^{2.3}$.

Following Edgar \& Chevalier (1986), we have used the cooling functions in
Sutherland \& Dopita (1993) to explicitly calculate the columns
for the most important ionic species that peak in fractional
abundance at $\log~T =$ 5.0 to 6.5 for
gas cooling radiatively
from an initial temperature $T = 5 \times 10^6$ K. We have considered
gas that cools at constant pressure (isobaric) and constant density
(isochoric). We further assume that radiative cooling by dust grains is not
significant (e.g. Draine 1981), that the
gas is optically thin to its cooling radiation, and that the relative
abundances of all the heavy elements have solar values. The results are
listed in Table 1.
\footnote{In this regime, greater care in discussing the metallicity dependence
of the cooling columns is required.
For the temperature range where the O{\sc iv}, O{\sc v}, and
O{\sc vi} ions are most abundant, Oxygen is the most important coolant.
Similarly, at temperatures where the Ne{\sc viii} ion is most abundant,
Neon is the most important coolant (Sutherland \& Dopita 1993).
Thus the relevant cooling time scales inversely with the O or Ne abundance,
implying that the ionic cooling columns are independent of the 
abundance.
In contrast, ions like 
Mg{\sc x}, Si{\sc xii}, O{\sc viii}, Ne{\sc ix}, and Ne{\sc x} reach peak
abundances at temperatures
where Fe is the most important coolant. Thus our estimates in Table 1
for cooling columns need to be multiplied by the ratio of the
actual elemental adundance relative to solar.
In the IGM, alpha-element/Fe abundance ratios
of up to twice solar are plausible (e.g.
Gibson, Loewenstein, \& Mushotzky 1997).}

First, consider the Li-like species
Ne{\sc viii}, Mg{\sc x}, and Si{\sc xii}
which have strong ultraviolet
resonance transitions analogous to the O{\sc vi} doublet
(Ne{\sc viii}$\lambda$$\lambda$770,780;
Mg{\sc x}$\lambda$$\lambda$ 610,625; 
Si{\sc xii}$\lambda$$\lambda$499,521). These transitions could be observed
with $FUSE$ in intergalactic clouds with suitable (moderate) redshifts,
and trace ions that reach their peak
abundances at $\log~T$ = 5.85 (Ne{\sc viii}) to 6.35 (Si{\sc xii}). 
In both the isobaric and isochoric calculations, these ions all have
cooling columns comparable to O{\sc vi}. Since these transitions
have similar oscillator strengths, they will have similar
optical depths and equivalent widths, 
and should be only slightly more difficult
to detect than the O{\sc vi} line. See Table 1 for details.
Note that the cooling columns for the Li-like species that trace cooler gas
(C{\sc iv} and N{\sc v})
are 10 to 50 times smaller than O{\sc vi}.

Next, consider the other abundant ionic species with strong resonance
lines in the EUV band. Notable examples are O{\sc iv}$\lambda$788,
O{\sc v}$\lambda$630, Ne{\sc v}$\lambda$568, and 
Ne{\sc vi}$\lambda$559. Again Table 1 shows that these species
have cooling columns comparable to O{\sc vi} (particularly for the
ioschoric models) and the implied
optical depths and equivalent widths are comparable
to the O{\sc vi}$\lambda$1032 line.

Finally, consider the Hydrogen-like and Helium-like species of
Oxygen and Neon. The cooling columns are very high ($\sim 10^{17}$
cm$^{-2}$) because of the steep temperature dependence noted above.
These ions have their resonance lines
at high energies, and
X-ray absorption studies are required to observe them.

To provide a graphical summary for a much wider array of
astrophyically-significant ions, we show in Figure 3 the cooling columns
estimated using the approximation
$N_{X(i)} = 3 k T_{X(i)}$ $\cal M$ $c_s (X/H)_{\odot} Z f_i /\Lambda$. This was
evaluated 
at the temperature where ionic species $i$ of element $X$ reaches
it peak abundance ($T_{X(i)}$), assuming $v_{cool} = c_s$
at $T_{X(i)}$ (i.e. $\cal M$ = 1). 
The values plotted are approximate
only, and are superceded by the detailed calculations described above
for the specific subset of ions in Table 1. The strong temperature
dependence of the cooling column is readily apparent in this figure.

\subsection{Application to the PKS 2155-304 Absorber}

Recent $Chandra$ and $FUSE$ data on the sight-line to the AGN
PKS 2155-304 enable us to directly test the above predictions. 
Nicastro et al. (2002) have used the Low-Energy Transmission
Grating Spectrograph on $Chandra$ to detect
X-ray absorption-lines due to H-like and He-like Oxygen and He-like Neon,
and they have compared these data to the $FUSE$ detection of
O{\sc vi} absorption (Savage et al. 2000; Sembach et al. 2000).

The $FUSE$ data have high enough spectral resolution to show that
the O{\sc vi} absorption arises in two components (a narrower
component centered at $v_{LSR}$ = +80 km s$^{-1}$ and a broader component
centered on $v_{LSR}$ = -92 km s$^{-1}$).
Nicastro et al. show that the relative column densities
of the O{\sc vi}, O{\sc vii}, O{\sc viii}, and Ne{\sc ix}
ions are inconsistent with single-temperature gas in collisional
ionization equilibrium. They find that a model of highly dilute gas
photoionized by the metagalactic background provides a satisfactory
fit to the observed columns provided that the Ne/O abundance ratio
is 2 to 3 times solar. This model requires that the absorber have a very low
density ($\sim 10^{-5}$ cm$^{-3}$) and correspondingly large size
(several Mpc). Thus, even though its small radial velocity suggests
that the absorber is part of the disk or halo of the Milky Way, its
immense implied size leads Nicastro et al. (2002) to propose that
it is a local intergalactic filament (part of the ``Warm-Hot Ionized
Medium'').

Here we propose an alternative interpretation. We suggest that the absorber
arises in gas that is cooling radiatively from an
initial temperature high enough to provide O{\sc viii} absorption, but
not Ne{\sc x} (e.g. $2 \times 10^6 K\leq T \leq 5 \times 10^6 K$).
This model has an important physical difference from the collisional
ionization equilibrium models ruled out by Nicastro et al.: while
they considered gas at a single temperature, the radiatively cooling
model contains gas that spans a broad temperature range.  

We can use the calculations described in the previous section to compare
to the Nicastro et al. data.
The results are listed in Table 2. There is a rough agreement, but the
models predict larger columns in the O{\sc vii}, O{\sc viii}, and Ne{\sc ix}
ions than derived from the X-ray data. Optical depth effects may play a role.
For cooling gas, the predicted optical depths in these X-ray lines are 
roughly 3 to 10 times higher than in the O{\sc vi}$\lambda$1032 line
(see Table 1).
If this is the case, then the X-ray lines in PKS 2155-304 could be
moderately optically-thick, and the
column densities reported in Nicastro et al. (2002) should then
be taken as lower limits. This would improve the agreement between
the models and the data. In fact,
the curve-of-growth analysis reported in Nicastro et al. 
shows that optical depth effects will be non-negligible in the
O{\sc vii} and Ne{\sc ix} lines if the Doppler
b-parameter is similar to that of the
narrow or broad O{\sc vi} absorbers (65 or 95 km s$^{-1}$
respectively).

If our model is correct, it has an attractive feature. Unlike the
photoionization model, it does not require 
the absorber to have a very low
density and very large size-scale. This gas could well
reside in the disk or halo of the Milky Way (as might be expected
based on its small LSR velocity).
 
\section{Conclusions}

Coronal-phase ($T \sim 10^5$ to $10^6$ K) gas is important in many
astrophysical contexts, but has been difficult to observe. The
O{\sc vi}$\lambda$$\lambda$1032,1038 doublet is a direct probe of this phase,
and has been recently detected in absorption by $FUSE$
in the disk and
halo of the Galaxy, high velocity clouds, the Magellanic Clouds, starburst
galaxies, and the IGM.

In this paper, we have presented a unified analysis of all these data.
We have shown that these disparate systems
define a simple, relatively tight relationship between the O{\sc vi}
column density ($N_{OVI}$) and absorption-line
width (b). The relationship is linear for broad lines (b $\geq$ 40 km s$^{-1}$)
but rolls over and steepens for narrower lines. The different
classes of O{\sc vi} systems tend to populate distinct regions of
the relationship, with Galactic disk clouds and starburst outflows
at the extrema. We have shown that the relationship is not due to
selection effects.
We have also demonstrated it is independent
of the Oxygen abundance over the range O/H = one-tenth to twice solar
but breaks down for the extremely metal-poor starburst galaxy
I~Zw~18 (O/H $\lesssim$ 2\% solar).
 
We have argued that the observed relation between O{\sc vi} column density
and line-width is exactly that
predicted theoretically as radiatively cooling hot gas passes
through the coronal temperature regime. The predicted column density 
is independent of the gas
density or metallicity (as long as the latter is $\gtrsim$
0.1 solar, so that the radiative cooling is dominated by Oxygen).
The O{\sc vi} column density depends only the characteristic flow
speed in the cooling gas (e.g. Edgar \& Chevalier 1986).
{\it This model (which has essentially no adjustable parameters)
naturally produces both the form and absolute
normalization of the observed column-density vs. line-width relation.}
It unifies systems spanning a broad range in size and dynamical state
ranging from very small (very large) quiescent clouds in the disk of the
Milky Way
(IGM), to meso-scale transonic flows in the halo of the Milky Way and Magellanic
Clouds, to supersonic global winds in some starbursts.

We infer that the intergalactic O{\sc vi} clouds can not have
metallicities less than a few percent solar, or they would lie
too far off the $N_{OVI}$ {\it vs.} b relation.
In order to be able to cool radiatively in less than a Hubble time,
the clouds must be smaller than $\sim$1 Mpc in size.
We have shown that if the O{\sc vi} clouds are collisionally-ionized,
the implied global
heating rate is consistent with available sources
(structure formation and/or galactic winds).
We have also predicted that column densities
comparable to those seen in O{\sc vi} ($\sim10^{14}$
cm$^{-2}$) will be
detected
in the other Li-like ions Ne{\sc viii}, Mg{\sc x}, and Si{\sc xii}
(if the intergalactic gas is cooling from $T \gtrsim 10^6$ K), 
and in the O{\sc iv}, O{\sc v}, Ne{\sc v}, and Ne{\sc vi} ions as well.

More generally, we have shown that the steep increase in radiative
cooling time with temperature means that the Li-like ions can be used
to probe radiatively cooling gas at temperatures up to an order-of-magnitude
higher than the value where the ionic fraction peaks (e.g. at up to
a few $\times 10^6$ K for O{\sc vi}).
We have also calculated radiative
cooling column densities for all abundant ionic species whose peak
fractional abundance occurs at $T \geq 10^5$ K. The H-like (He-like)
O, Ne, Mg, Si, and S
ions have characteristic cooling columns of $\sim10^{17}$ (few$\times10^{16}$)
cm$^{-2}$, but must be observed with X-ray telescopes.

Finally, we have shown that the
properties of the O{\sc vii}, O{\sc viii}, and Ne{\sc ix} absorption-lines
recently detected with $Chandra$ (Nicastro et al. 2002) 
may be consistent with a scenario of radiatively
cooling gas in the Galactic disk or halo.



\acknowledgments

The authors thank Robert Benjamin, Ed Jenkins, and Blair Savage for
useful suggestions. We also thank Alessandra Aloisi, Charles Hoopes,
Chris Howk, and Crystal Martin for their assistance in the analysis of some
of the $FUSE$ data shown in this paper. We thank an anonymous referee
for several suggestions that improved the paper.
This work was supported in part by NASA FUSE GI grants NAG5-9012
and NAG5-10302.
TMH and KRS also
acknowledge support from NASA Long-Term Space Astrophysics grants
NAG5-6400 and  NAG5-3485 respectively. DKS is
supported by NASA through Chandra Postdoctoral Fellowship Award PF0-10012, 
issued by the Chandra X-ray Observatory Center, which is operated by the
Smithsonian Astrophysical Observatory for and on behalf of NASA under
contract NAS8-39073.



\appendix

\section{The Temperature Dependence of Cooling Column for Li-like Ions}

In Figure 4 we show the fractional abundance of O{\sc vi} ions
as a function of temperature,
assuming collisional ionization equilibrium (taken from Sutherland
\& Dopita 1993). The asymmetric ``tail'' on the high-temperature
side of the peak at $logT \sim$ 5.5 is a consequence of the broad temperature
range over which O{\sc vii} (from which O{\sc vi} is produced
by recombination) is the dominant ion.

In Figure 5 we plot the temperature dependence of the value of the 
O{\sc vi} cooling column density
predicted from $N_{OVI} = (3 kT {\cal M} c_s/\Lambda(T,Z)   
f_{OVI}(T) O/H$ (taking the flow's Mach number $\cal M$ = 1). The maximum
value of the column density provides a reasonable approximation 
to the total column density derived from a more
detailed calculation, 
where the cooling of a parcel of gas is followed in detail as a 
function of time. 

The secondary peak at $logT \approx$ 6.4, is a consequence of the
O{\sc vi} fractional abundance remaining significant up to this temperature
(Figure 4).
The total gas cooling column scales with the cooling time (which increases
as $\sim T^{1.8}$) and with the flow speed (which increases
with $T^{0.5}$ for constant $\cal M$). These effects compensate for the
decrease in the 
ionic fraction of O{\sc vi} with increasing temperature. 

Figure 5 therefore shows that for appropriately high initial temperatures,
a significant fraction
of the O{\sc vi} column density in radiatively cooling gas
can arise in gas at temperatures up to an order-of-magnitude higher
than simple consideration of collisional ionization equilibrium would suggest.
For the case of a flow with constant Mach number, the fractional contribution
to $N_{cool}$ from gas hotter than $10^6$ K is about 45\%.

Note that above $log~T \sim$ 6.0 the primary coolant is Fe, and our
calculations assume a solar O/Fe ratio. For the more general case,
the predicted cooling columns at $T \geq 10^6$ K should be multiplied by
the actual O/Fe relative to the solar value.

Similar arguments apply to the other Li-like species discussed above
(e.g. the Mg{\sc x} ion could be used to probe radiatively cooling gas
at temperatures up to almost $10^7$ K).




\clearpage


\newpage
\clearpage

\clearpage
\begin{figure}[ht!]
\plotone{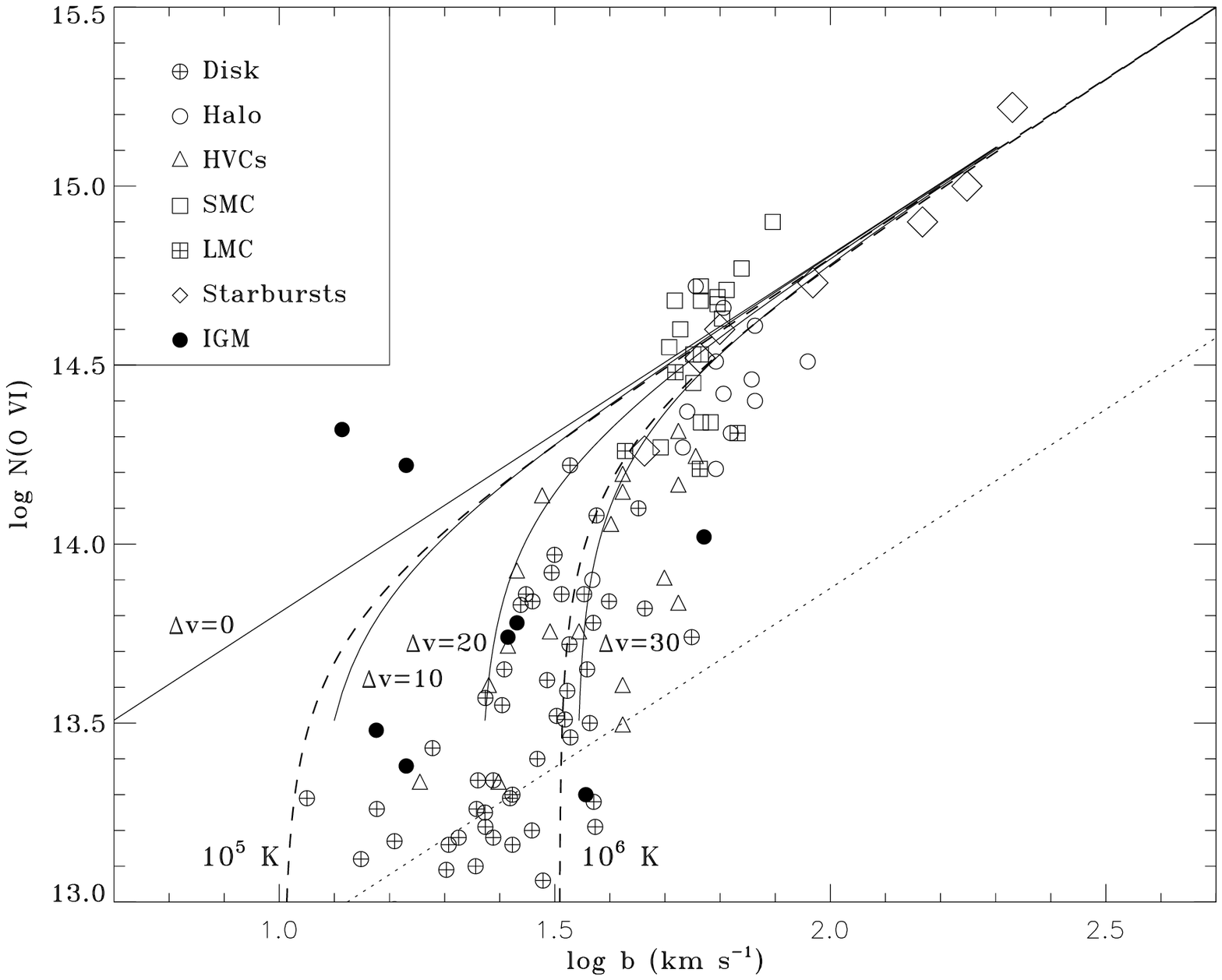}
\caption{Column density versus line width for \ion{O}{6} absorption
systems observed in the Galactic disk and halo, high velocity clouds, the 
Magellanic Clouds (LMC, SMC), starburst galaxies, and the intergalactic 
medium (IGM).  Typical errors on these data points are 0.05 to 0.10 dex 
for $\log N_{OVI}$ and 0.05 to 0.08 dex for $\log \sigma$.  
The relation predicted for radiatively cooling gas is shown as the
two long-dashed lines for an assumed temperature $T_{OVI} = 10^5$ K
and $10^6$ K respectively. The short-dashed line in the lower right
corner represents the typical detection limit in the $FUSE$ spectra.
The solid lines show the effects of the blending of multiple narrow
components with velocity separations as indicated. See text for details.
\label{fig1}}
\end{figure}

\begin{figure*}
\plotone{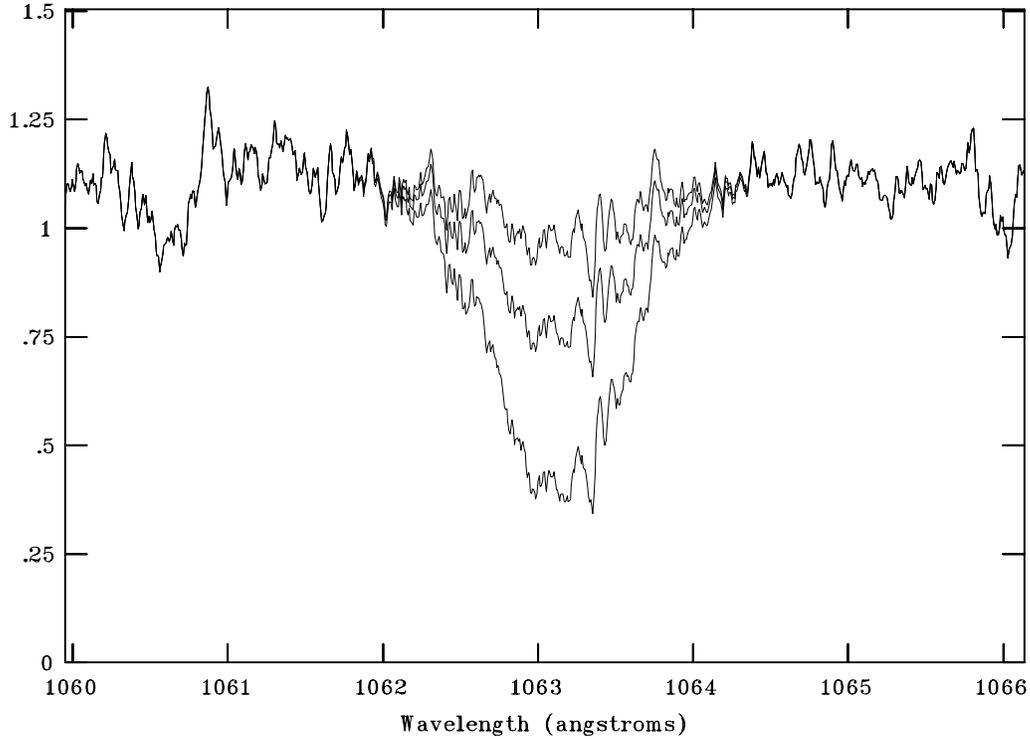}
\caption{Spectra showing the addition of three broad (log b = 2.25) redshifted
O{\sc vi}$\lambda$1031.9 absorption-lines to a $FUSE$ spectrum.
From top to bottom these
have residual intensities at line center of 85\%, 64\%, and 28\%.
The strongest line corresponds to the typical broad O{\sc vi}
lines (log b $>$ 2.0) in Figure 1. The weakest line corresponds to
our adopted $FUSE$ detection threshold (dashed diagonal line in Figure 1).
See text for details.}
\label{fig2}
\end{figure*}

\begin{figure*}[!hb]
\plotone{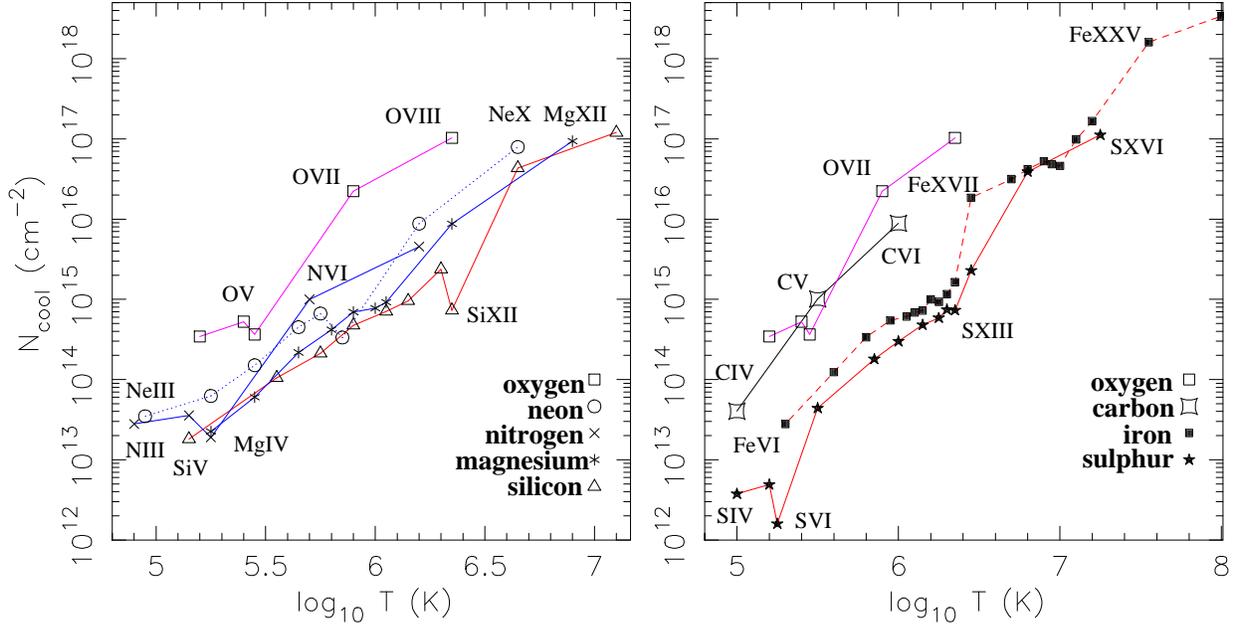}
\caption{Approximate cooling column densities for various ions of abundant
elements, plotted at the temperature at which each ion has its maximum
fractional abundance assuming collisional ionization equilibrium.
The cooling columns are proportional to the Mach number of the flow,
and are plotted for $\cal M$ = 1.
To maximize the clarity of the figure, the data are presented in two
separate panels, and only some of the ions are explicitly identified.
Oxygen cooling columns are plotted in both panels to aid comparison
between the two panels. This figure is illustrative only, and more accurate
cooling columns for selected ions based on detailed calculations
are listed in Table 1. See text for details.
\label{fig3}}
\end{figure*}

\begin{figure*}
\plotone{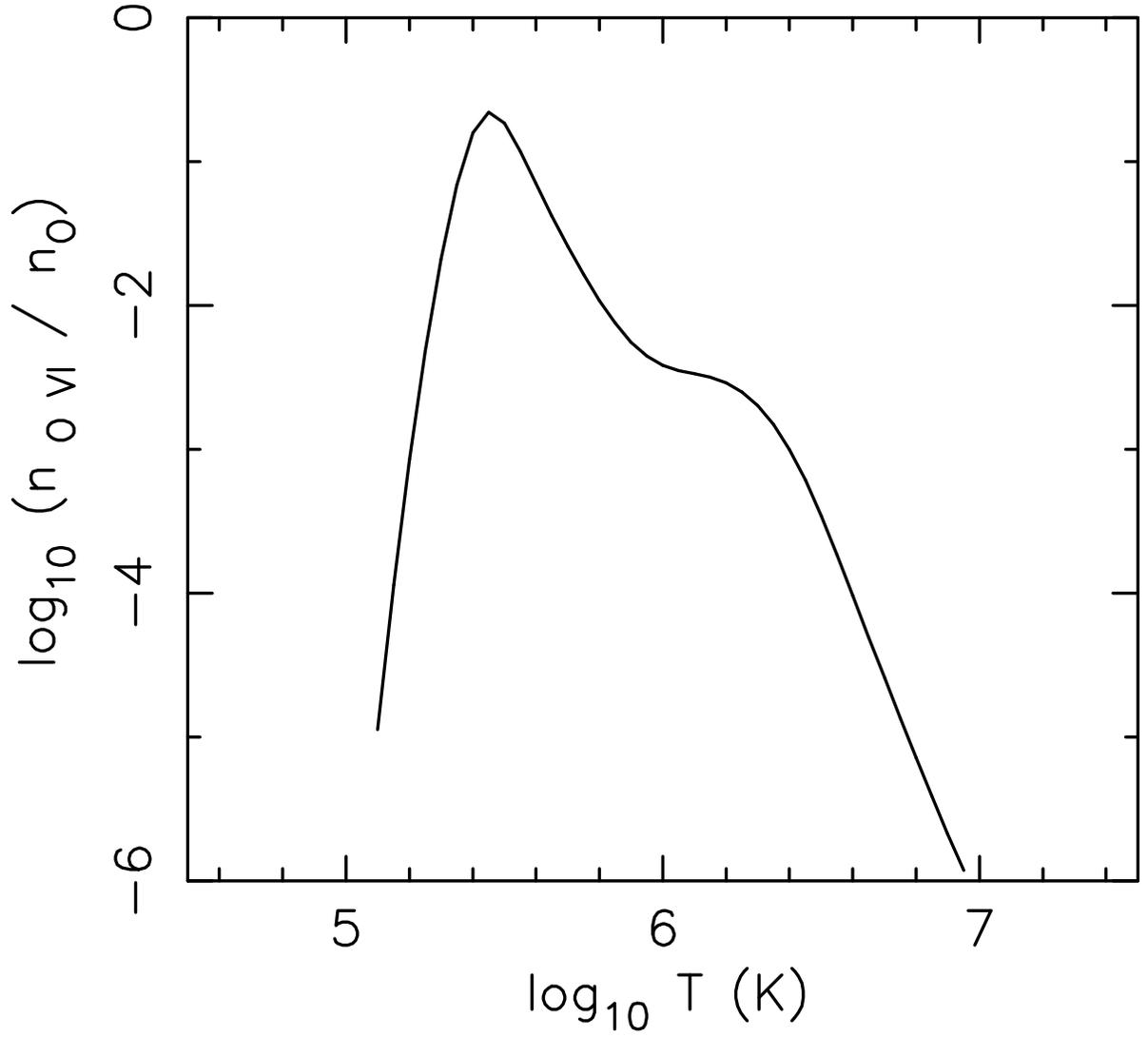}
\caption{Fractional abundance of the O{\sc vi} ion as a function of
temperature, assuming collisional ionization equilibrium (Sutherland \&
Dopita 1993).
\label{fig4}}
\end{figure*}

\begin{figure*}
\plotone{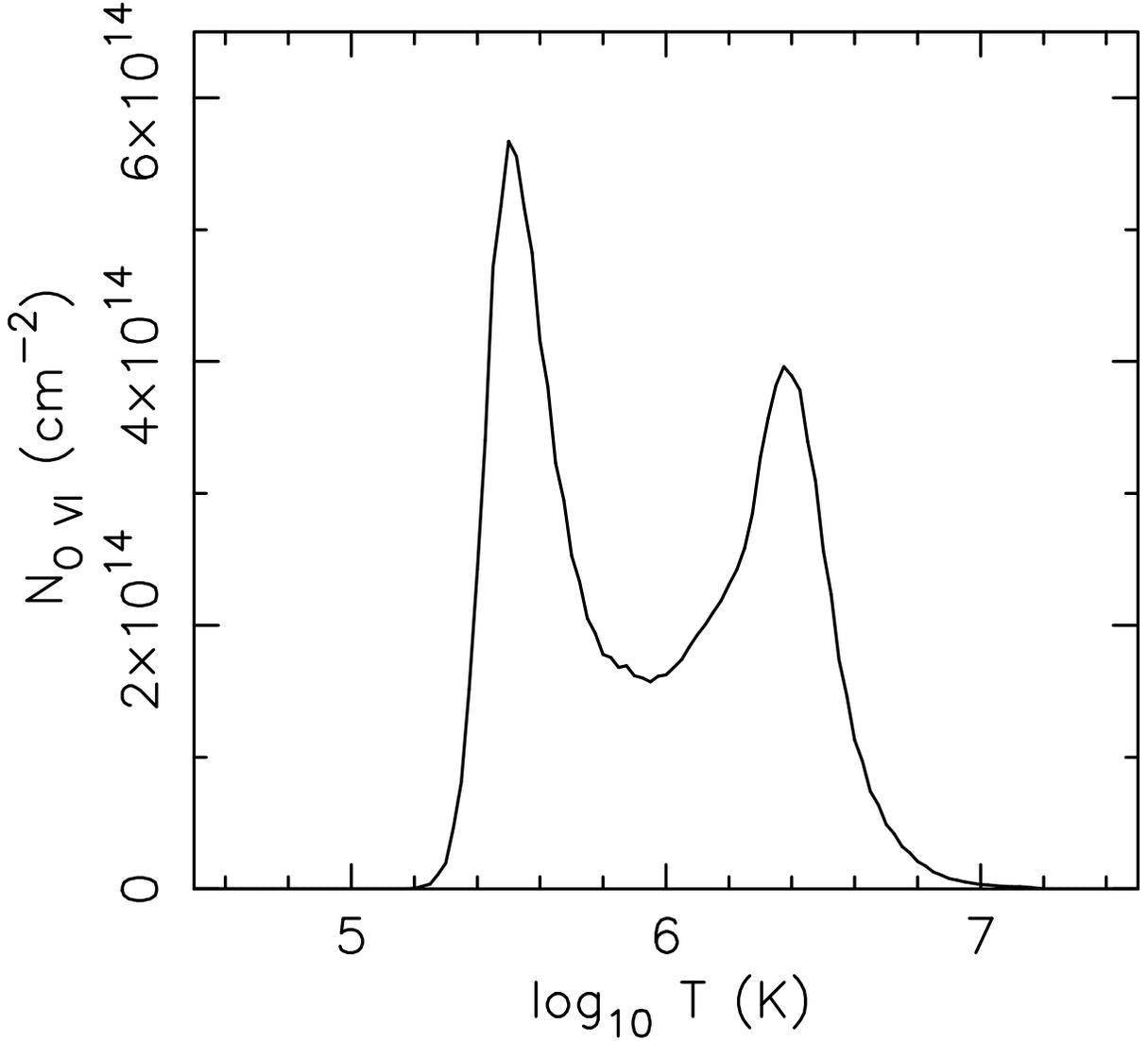}
\caption{Predicted
column density for the O{\sc vi} ion as a function of
temperature for gas cooling radiatively from $T \geq 10^7$ K at
a velocity $v_{cool} = c_s$ ($\cal M$ = 1). The
secondary peak at high temperature is primarily a consequence
of the increase in radiative cooling time with temperature,
together with the moderate O{\sc vi} ionic fraction at
temperatures up to a few million degrees (Figure 4). 
See text for details.
\label{fig5}}
\end{figure*}



\clearpage

\begin{deluxetable}{lccccccc}
\tablecolumns{0}
\tablewidth{0pt}
\tablecaption{Selected Cooling Columns and Absorption-Line Strengths}
\tablehead{Ion & $log~T_{max}$ & $log N_{ib}$ & $log N_{ic}$ & Line & $log~N_{ib} f \lambda$ & $log~N_{ic} f \lambda$ & $\tau$$_{ic}$}
\startdata
C{\sc iv} & 5.00 & 12.5 & 13.9 & 1548.2 & 7.0 & 8.4 & 0.25\\
N{\sc v} & 5.25 & 12.6 & 13.6 & 1238.8 & 6.9 & 7.9 & 0.08\\
O{\sc iv} & 5.20 & 13.6 & 15.0 & 787.7 & 7.5 & 8.9 & 0.8\\
O{\sc v} & 5.40 & 13.6 & 14.8 & 629.7 & 8.1 & 9.3 & 2.0\\
O{\sc vi} & 5.45 & 14.2 & 15.0 & 1031.9 & 8.4 & 9.1 & 1.3\\
O{\sc vii} & 5.90 & 16.4 & 16.9 & 21.60 & 9.6 & 10.1 & 13\\
O{\sc viii} & 6.35 & 16.8 & 17.0 & 18.97 & 9.5 & 9.7 & 5\\
Ne{\sc v} & 5.45 & 13.4 & 14.6 & 568.4 & 7.2 & 8.4 & 0.25\\
Ne{\sc vi} & 5.65 & 13.7 & 14.7 & 558.6 & 7.6 & 8.6 & 0.4\\
Ne{\sc viii} & 5.85 & 14.3 & 14.7 & 770.4 & 8.2 & 8.6 & 0.4\\
Ne{\sc ix} & 6.20 & 16.3 & 16.5 & 13.45 & 9.3 & 9.5 & 3\\
Mg{\sc x} & 6.05 & 14.5 & 14.8 & 609.8 & 8.2 & 8.5 & 0.3\\
Si{\sc xii} & 6.35 & 14.8 & 15.0 & 499.5 & 8.3 & 8.5 & 0.3\\
\enddata
\tablecomments{The temperature at which each ion reaches its peak 
abundance in CIE is given in column 2. The calculations of cooling columns
are based on the cooling flow behind a 600 km s$^{-1}$ shock (postshock
temperature $5 \times 10^6$ K) using the Sutherland \& Dopita (1993)
non-equilibrium zero-field cooling function and solar abundances. They assume
that the cooling proceeds under isobaric conditions
(column 3) or isochoric conditions (column 4). For the specific
resonance absorption-lines whose wavelength is listed in column 5
(in \AA), the implied line strengths
for the isobaric and isochoric models are given in columns 6 and 7
respectively. These are given in terms of the product of the column
density, oscillator strength (htpp://physics.nist.gov/), and
wavelength (in cm). The optical depth at line center
is just $\tau = 0.015 N f \lambda/ b$ (where the Doppler parameter
b is in cm s$^{-1}$). The implied optical depths at line-center for
the isochoric model (b = $v_{shock}/4$ = 150 km s$^{-1}$) are given in column 8.
The equivalent widths in the optically-thin
limit are $W/\lambda = 8.85 \times 10^{-13} N f \lambda$ (e.g Spitzer 1978).}
\end{deluxetable}

\clearpage

\begin{deluxetable}{lcccc}
\tablecolumns{0}
\tablewidth{0pt}
\tablecaption{Radiative Cooling Models for the PKS 2155-304 Absorber}
\tablehead{Columns & Data (both) & Data (single) & Isobaric & Isochoric}
\startdata
log~$N_{OVI}$ & 14.4$\pm$0.1 & 14.0$\pm$0.1 & 14.2 & 15.0\\
log[$N_{OVII}/N_{OVI}$] & $\geq$1.2$\pm$0.3 & $\geq$1.6$\pm$0.3 & 2.2 & 1.9\\
log[$N_{OVIII}/N_{OVI}$] & $\geq$1.3$\pm$0.4 & $\geq$1.7$\pm$0.4 & 2.6 & 2.0\\
log[$N_{NeIX}/N_{OVI}$] & $\geq$1.4$\pm$0.3 & $\geq$1.8$\pm$0.3 & 2.1 & 1.5\\
\enddata
\tablecomments{The measured values for the column densities are listed
twice. In column 2 we assume both the broad and narrow
O{\sc vi} components seen in the $FUSE$ spectrum are associated with the
X-ray absorber.
In column 3, we assume that only the broad O{\sc vi} component
is associated with the X-ray absorber (as proposed by
Nicastro et al. 2002). In columns 2 and 3 the values for $N_{OVII}$,
$N_{OVIII}$, and $N_{NeIX}$ are treated as lower limits to allow for
the possibility that the X-ray absorption-lines are not optically-thin.
The values in columns 4
and 5 for isobaric and isochoric cooling are taken from the same
calculations
reported in Table 1.}
\end{deluxetable}


\end{document}